\pdfoutput=1
\documentclass[prx,twocolumn,amsmath,amssymb,eqsecnum,nofootinbib,superscriptaddress]{revtex4-2}
\usepackage{graphicx,amsmath,relsize,epstopdf,color,mathtools,bm,newtxtext,newtxmath,braket,rotating}
\usepackage[hyphenbreaks]{breakurl}
\usepackage[colorlinks=true,linkcolor=blue,citecolor=blue,urlcolor =blue]{hyperref}
\usepackage[normalem]{ulem}

\usepackage{booktabs}

\usepackage[table,xcdraw]{xcolor}
\newcommand{\eq}[1]{\begin{equation}\begin{aligned}#1\end{aligned}\end{equation}}

\newcommand{\eu}{\mathrm{e}}
\newcommand{\iu}{\mathrm{i}}
\newcommand{\ha}{\hat{a}}
\newcommand{\had}{\hat{a}^\dagger}
\newcommand{\hb}{\hat{b}}

\newcommand{\sugg}[1]{#1}
\newcommand{\vac}{\ket{\mathrm{vac}}}

\usepackage{soul,xcolor}

\usepackage{dsfont}

\begin{document}
	
\setstcolor{red}

\title{Teleamplification on the Borealis boson-sampling device}

\begin{abstract}
    A recent theoretical proposal for teleamplification requires preparation of Fock states, programmable interferometers, and photon-number resolving detectors to herald the teleamplification of an input state. {These enable teleportation and heralded noiseless linear amplification of a photonic state up to an arbitrarily large energy cutoff.} We report on adapting this proposal for Borealis and demonstrating teleamplification of squeezed-vacuum states with variable amplification factors. 
    The results match the theoretical predictions and exhibit features of amplification in the teleported mode, with fidelities from 50-93\%. This demonstration motivates the continued development of photonic quantum computing hardware for noiseless linear amplification's applications across quantum communication, sensing, and error correction.
\end{abstract}

\author{Aaron Z. Goldberg}
\affiliation{National Research Council of Canada, 100 Sussex Drive, Ottawa, Ontario K1N 5A2, Canada}
\affiliation{Department of Physics, University of Ottawa, Advanced Research Complex, 25 Templeton Street, Ottawa, Ontario K1N 6N5, Canada}

\author{Khabat Heshami}
\affiliation{National Research Council of Canada, 100 Sussex Drive, Ottawa, Ontario K1N 5A2, Canada}
\affiliation{Department of Physics, University of Ottawa, Advanced Research Complex, 25 Templeton Street, Ottawa, Ontario K1N 6N5, Canada}
\affiliation{Institute for Quantum Science and Technology, Department of Physics and Astronomy, University of Calgary, Alberta T2N 1N4, Canada}
\maketitle
%\tableofcontents

\section{Introduction}
Loss is a serious challenge to photonic quantum information processing. Given the impossibility of noiselessly amplifying arbitrary quantum states \cite{Louiselletal1961,HausMullen1962,Heffner1962,Gordonetal1963,Caves1982,Clerketal2010,Cavesetal2012}, loss must only be avoided. Noiseless amplification, however, does exist in specific scenarios: any state with a maximal photon number can be perfectly amplified using a recent \textit{probabilistic} protocol \cite{Guanzonetal2022}. In fact, this protocol can herald the perfect amplification of any state up to a resource-dependent cutoff, while simultaneously achieving quantum-state teleportation. We here demonstrate this ``teleamplification'' procedure on Xanadu's machine Borealis.

Noiseless deterministic amplification would violate the no-cloning theorem \cite{Wigner1997,Park1970,WoottersZurek1982,Dieks1982,GhirardiWeber1983} by allowing the transformation $\ket{\alpha}\to\ket{\sqrt{2}\alpha}$ followed by a beam-splitter transformation $\ket{\sqrt{2}\alpha}\to\ket{\alpha}\otimes\ket{\alpha}$ for any input coherent state $\ket{\alpha}$, which would solve numerous problems such as transportation of quantum states over large distances \cite{Yangetal2013}, while rendering quantum key distribution schemes insecure \cite{GrosshansGrangier2002} and allowing for superluminal communication \cite{Herbert1982,Gisin1998}. Quantum theory allows two alternatives: noise-added amplification, which places an upper bound on the fidelity between the actual amplified state and the desired amplified state \cite{Lindblad2000,Cerfetal2000,CerfIblisdir2000,GrosshansGrangier2001,Fiurasek2001,Braunsteinetal2001,Faseletal2002,Andersenetal2005,Josseetal2006}, and nondeterministic amplification, which may achieve perfect fidelities at the expense of not always succeeding \cite{RalphLund2009,Xiangetal2010,Usugaetal2010,Ferreyroletal2010,Ferreyroletal2011,Zavattaetal2011,NeergaardNielsenetal2013,Hawetal2016,Parketal2016,ZhangZhang2018,Guanzonetal2022arxiv}. A new scheme of the latter type was recently developed \cite{Guanzonetal2022}, with the benefits of being heralded, such that one can be certain which of a series of trials succeeded at amplification, of enjoying significantly higher success probabilities than previous methods, and of working for all possible input states. The scheme requires a fixed linear optical network with one programmable beam-splitter transmissivity to control the amount of amplification, supplemented with either single-photon input states and photon-number-resolving detectors (PNRDs) or a Fock-state input and detectors capable of distinguishing between zero, one, and more than one photon at the output. Then, any arbitrary input state can be fed into the circuit for probabilistic, heralded amplification with arbitrarily large fidelity increasing with the number of modes in the fixed circuit.

\sugg{Teleamplification, being an ideal form of noiseless linear amplification, has applications across quantum information science. These include quantum key distribution \cite{Gisinetal2010,Blandinoetal2012}, quantum error correction \cite{Ralph2011}, imaging beyond the diffraction limit \cite{KellererRibak2016},  and repeaters for quantum communication \cite{Zhaoetal2017,Zhouetal2020,Diasetal2020}. In a recent work, it was demonstrated that a form of noiseless linear amplification using a continuous-variable Bell state can improve the decoherence resistance of teleportation protocols \cite{Zhaoetal2023}. Amplification using only linear interferometers was already demonstrated in the noise-added case in Ref. \cite{Josseetal2006}, while linear interferometers and postselection fruitfully combine for a number of applications \cite{Lassenetal2010,Matthewsetal2011,Motesetal2015,Matthewsetal2016,Tzitrinetal2020GKP,Thekkadathetal2020}. We here go beyond Ref. \cite{Zhaoetal2023}'s demonstration of teleportation and amplification of coherent and displaced thermal states by teleamplifying squeezed states, use a different protocol without nonlinear elements, and attain a maximum fidelity of 92.6(9)$\%$ to rival Ref. \cite{Zhaoetal2023}'s 92$\%$ when comparing the same gain factor of $g=1$.
}

Gaussian boson sampling has the same requirements as the teleamplification protocol: programmable interferometers and PNRDs. The only difference is that Gaussian boson sampling requires squeezed light as input states while teleamplification requires Fock-state inputs, but Fock states can be generated by heralding the detection of $n$ photons in one arm of a two-mode squeezed vacuum state (TMSV). Teleamplification is thus amenable to Borealis, a Gaussian boson sampling device available on the cloud. 

A strong advantage of Borealis is its tunability. Borealis has been used to demonstrate quantum advantages in boson sampling \cite{Madsenetal2022}, as well as for measuring quantumness quantifiers \cite{Goldbergetal2023QCS} and solving graph theory problems \cite{Stanevetal2023arxiv}. Instead of having to construct a bespoke teleamplification device, Borealis can readily be adapted to the task at hand, and we used it to test teleamplification of a variety of $P$-nonclassical input states with a variety of amplification factors. These together demonstrate the usefulness of Gaussian boson sampling devices throughout quantum information processing applications.

\section{Methods}

\subsection{\sugg{Teleamplification's goal}}
We begin with a\sugg{n arbitrary single-mode} quantum state expressed in the Fock basis as
\eq{
\ket{\psi}=\sum_{k=0}^\infty\psi_k\ket{k},
} where the Fock states are created by bosonic operators ($[\ha,\had]=1$) acting on the vacuum $\ket{k}=\ha^{\dagger k}\vac/\sqrt{k!}$. One such state is the coherent state $\ket{\alpha}$ with a Poissonian photon-number distribution $\psi_k\propto \alpha^k/\sqrt{k!}$. Ideal noiseless amplification up to a linear cutoff enacts
\eq{
\ket{\psi}\underset{g;n}{\to}\ket{g\psi;n}\propto\sum_{k=0}^n g^k\psi_k\ket{k}.
\label{eq:teleamp}
} In the limit of $n\to \infty$ or for states with a maximum initial photon number, this amplification is perfect $\ket{g\psi;n}\propto g^{\had\ha}\ket{\psi}$. Such a transformation magnifies all coherent-state amplitudes as $\alpha\to g\alpha$ regardless of their initial phases \sugg{and could be used to perfectly clone any input coherent state using the method given in the introduction.}. \sugg{Noise-added amplification using linear optics and tunable $g$ was demonstrated in Ref. \cite{Sabuncuetal2007}.} \sugg{In contrast, since teleamplification} can be made as ideal as desired, it must be probabilistic.

\subsection{\sugg{Teleamplification ingredients}}
To implement Eq.~\eqref{eq:teleamp}, Ref. \cite{Guanzonetal2022} proposed to combine two ingredients: (1) a variable beam splitter with transmissivity $\tau=g^2/(1+g^2)$ acting on the vacuum and a Fock state $\ket{n}$ and (2) a linear optical circuit that implements a Fourier transform on $n+1$ modes of which $n-1$ begin in the vacuum state, one ends in the vacuum state, and \sugg{the} other \sugg{$n$} end in single-photon states. Putting these two together yields an operator proportional to $\sum_{k=0}^n g^k\ket{k}\bra{k}$ on the appropriate mode. \sugg{There is a fixed set of beam splitters and phase shifters that are known to generate these specific linear optical circuits for any positive integer $n$; the Fourier transform circuit \cite{Zukowskietal1997,VourdasDunningham2005}, for example, has been demonstrated with four photons \cite{Suetal2017Fourier} and eight modes \cite{Crespietal2016}.}

 The state $\ket{\psi}$ and the output from the variable beam splitter are the two other inputs to the Fourier circuit, while the state $\ket{g\psi;n}$ gets teleported to the remaining output of the variable beam splitter. \sugg{To repeat, this teleportation and amplification is only successful when $n$ of the Fourier transform's outputs are measured to be single photons and the other one to be the vacuum state.} Different patterns of where the vacuum state appears among the $n+1$ output modes yield teleamplified states with different rephasings of the coefficients $\psi_k$, which can be deterministically corrected just like in qubit teleportation where different Bell-state measurements require different updates to be performed on the state, parametrized by classical bits. 

\subsection{\sugg{Adapting the scheme for Borealis}}
\begin{figure*}
    \centering
    \includegraphics[width=\textwidth]{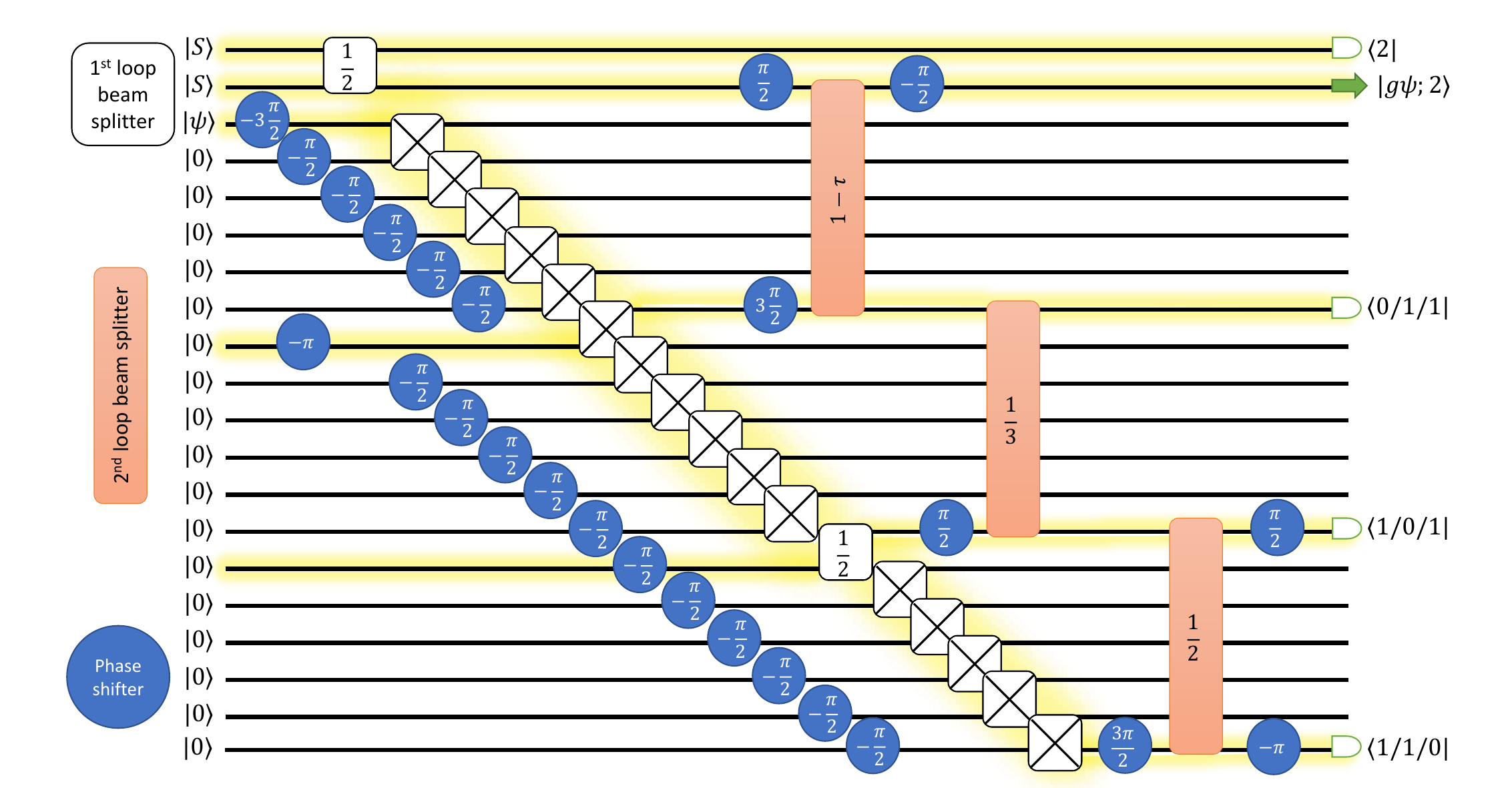}
    \caption{Circuit programmed for Borealis. There are 20 time bin modes unravelled from top to bottom for visualization purposes; \sugg{highlighted (in yellow) to display the flow of information in a simplified circuit are} the relevant five modes required for creating a Fock state (two) and implementing teleamplification up until $n=2$ ($|\psi\rangle$ and two vacuum modes). The first two modes have SMSV states such that the first is used to herald a two-photon Fock state in the second. The third mode has the state $\ket{\psi}$ to be amplified up to $n=2$, which can be any state and which is here another SMSV. The amount of amplification is set by the transmission parameter $\tau$. On Borealis, modes can interact with adjacent time bins by waiting in the first delay loop until the neighbour arrives, time bins six modes away by waiting at the second delay loop until the sixth-nearest neighbour arrives, and similarly for the 36th nearest neighbour. This is why all of the beam splitters in the sketch connect neighbours or sixth-nearest neighbours and cascade from top left to bottom right. We use rectangles for beam splitters with their probabilities \sugg{of being transmitted into the loop} written on top, with the X boxes \sugg{physically} implying complete reflection \sugg{and computationally implying complete transmission: whatever is in the loop is saved in the loop and whatever is outside the loop bypasses the loop so that each is directed to where the other would have gone if the beam splitter was absent}. The circles are phase shifts that are applied to a mode before it enters a delay loop.
    Most of the phases are present for the purpose of offsetting an automatic phase implemented in all of Borealis's beam splitters \sugg{and simply make the overall unitary operator given in Appendix~\ref{sec:validating transfer matrix} have mostly real entries}. The $3\pi/2$ phase in the final mode is necessary for the quantum Fourier transform. Extra phases of $\pi$ will be present sporadically due to the lack of full programmability of Borealis, so we simply ensure the compiled program has the correct transfer matrix, but many of them are actually irrelevant to the overall physics (such as phases on vacuum input states). Any of the postselection patterns (separated by ``/'') can be selected, which simply rephase the coefficients of $|g\psi;2\rangle$.  }
    \label{fig:circuit diagram unravel}
\end{figure*}

To enact this transformation on Borealis,\footnote{All source code for simulating and executing our protocols on Borealis as well as the measurement data and device certificates are available via Github at \url{https://github.com/AaronGoldberg9/Teleamplification_Borealis/}.} we first need access to a Fock state $\ket{n}$. Borealis carves from a single continuous-wave laser source at 775
nm a series of \sugg{pumps that produce} phase-stable single-mode squeezed vacuum states (SMSVs)
\eq{
\ket{\mathrm{SMVS}}=\frac{1}{\sqrt{\cosh r}}\sum_{k=0}^\infty (-\tanh{r})^k\frac{\sqrt{(2k)!}}{2^k k!}\ket{2k}
}
in time bins separated by $1/6\,\mathrm{MHz}$ with average power 3.7 mW and duration 3 ns per pulse, sends them through a series of phase elements and variably coupled delay lines such that many different time-bin modes have the opportunity to interact with each other, then sends each mode to one of an array of PNRDs to have its photon number measured. Further technical specifications and diagrams of the setup
can be found in Ref. \cite{Madsenetal2022}.
To make a Fock state with these resources,
we begin by combining two equal-magnitude SMSVs on a symmetric beam splitter, setting the phase on the beam splitter in such a way \cite{GoldbergHeshami2021} that they combine into a TMSV
\eq{
\ket{\mathrm{TMSV}}=\frac{1}{\cosh r}\sum_{k=0}^\infty (-\tanh{r})^k\ket{k}\otimes \ket{k}.
} We send one branch of this state to a PNRD, whose measurement result is an integer $n$ that heralds the presence of a Fock state $\ket{n}$ in the other branch. The state that we choose to amplify is also a SMSV, such that $n=2$ is the smallest nontrivial cutoff for teleamplification (note that $\ket{\mathrm{SMSV}}$ has no single-photon component). \sugg{This step is depicted in the top two rails of Fig.~\ref{fig:circuit diagram unravel}.}

The rest of the linear optical network is implemented using Borealis's programmable phase shifters and beam splitters before all of the modes are sent to PNRDs: some modes are used for heralding the success of the Fourier transform circuit and thus of the teleamplification, while one mode is measured to characterize the properties of the teleamplified state. The one caveat is that not every time-bin mode can directly interact with each other; the first delay loop allows neighbours to interact, the second pairs modes that are $6$ time bins apart, and the final brings together modes that are 36 time bins apart. For the appropriate five modes to interact (two to create the TMSV, one SMSV to be amplified, and $n=2$ additional vacuum states input to the Fourier transform circuit), we must extend our circuit to act over 20 modes of which the majority begin and end as vacuum states.

Our overall circuit diagram can be found in Fig.~\ref{fig:circuit diagram unravel}. This is an unravelled view of a time-bin circuit, with descending wires representing subsequent time bins. \sugg{Highlighted in yellow are the five rails that are necessary for the circuit to operate, while the rest of the diagram depicts modes that are necessary due to Borealis's specific architecture. A version of the diagram that focuses only on these highlighted rails can be found in Appendix~\ref{sec:simplified circuit diagram}.}

First, we perform the symmetric beam splitter on the two initial SMSV states to create TMSV by delaying the first mode at the first delay line to interact with the second mode when it arrives at the delay, then send the first time-bin mode to have its photon number measured. If the result of this measurement is $n=2$, we the heralding is successful and teleamplification may proceed. The second mode is sent to wait in the second delay line to perform more of the desired circuit. After waiting one delay period, it interacts with a vacuum mode six time bins away with a transmissivity $1-\tau$ that encodes the gain parameter, such that the second time bin will house the teleamplified state after the $2+6=8$th time-bin mode successfully completes the rest of the circuit. \sugg{Note that the ``$1-\tau$'' beam splitter is elongated in Fig.~\ref{fig:circuit diagram unravel} so that it connects two rails that are six rails apart.}

Next, since the SMSV states on Borealis are always made in adjacent time bins, we repeatedly delay the third time-bin mode that houses our state-to-be-amplified $\ket{\psi}=\ket{\mathrm{SMSV}}$ by transmitting it into the delay loop, reflecting it to stay in the loop, and reflecting onward all of the incoming vacuum states. One of these reflected vacuum states is the one whose interaction with the second time-bin mode we just described\sugg{, beginning in the 9th rail from the top of Fig.~\ref{fig:circuit diagram unravel}, reflecting into the 8th rail at the first beam splitter before approaching the $1-\tau$ beam splitter}. The delayed state $\ket{\psi}$ interacts with a vacuum mode at a balanced beam splitter \sugg{that is the topmost ``$X$'' box of Fig.~\ref{fig:circuit diagram unravel}} before half of it can finally exit the first delay stage; it goes on to interact with the other output from the (2,8) interaction described previously, now with the $8$th time-bin mode proceeding toward measurement and the 14th mode continuing through the circuit. The other half of $\ket{\psi}$ that was at the balanced beam splitter is delayed even further in the first delay loop until it is ready to interact with the $14$th mode as it exits the second delay loop. The modes are then sent to PNRDs to be measured. Throughout, relative phases are crucial: some phases are added to adjust the native beam-splitter phases implemented by Borealis, while the relative phase in between the first and second delay lines for the final mode is imperative to the Fourier transform circuit. Additional phases are always added during the delay loops that are compensated for by the Borealis compiler to the best of its ability.

\subsection{\sugg{Schematic verification}}
We can verify using the passive compiler attuned to Borealis that the entire circuit enacts
\eq{
U^\dagger \mathbf{a}U=\mathbf{U}\mathbf{a}
} for a particular transfer matrix $\mathbf{U}$, where we collect the mode operators in a vector $\mathbf{a}=(a_1,\cdots,a_{20})^\top$. This means that the state will evolve as $f(\mathbf{a}^\dagger)\vac\to f(\mathbf{U}^\top\mathbf{a}^\dagger)\vac$. 
Since only the first three modes begin with photons, we only need consider the first three columns of $\mathbf{U}$. Our circuit is composed in such a way that the only nonzero rows for these first three columns correspond to the first, second, eighth, fourteenth, and twentieth time bins (note the spacings by one or six due to the pertinent delay loops and see the corresponding highlighted portions of Fig.~\ref{fig:circuit diagram unravel}). We thus need to consider only the submatrix
\begin{equation}
    \mathbf{M}=\begin{pmatrix}
        \frac{1}{\sqrt{2}}&\frac{\iu}{\sqrt{2}}&0\\
        \iu\sqrt{\frac{1-\tau}{2}}&\sqrt{\frac{1-\tau}{2}}&0\\
        -\iu\sqrt{\frac{\tau}{6}}&-\sqrt{\frac{\tau}{6}}&-\frac{1}{\sqrt{3}}\\
        \iu\sqrt{\frac{\tau}{6}}&\sqrt{\frac{\tau}{6}}&\frac{1}{\sqrt{3}}\omega^2\\
        -\iu\sqrt{\frac{\tau}{6}}&-\sqrt{\frac{\tau}{6}}&-\frac{1}{\sqrt{3}}\omega       
    \end{pmatrix},
\end{equation} with $\omega=\exp(2\pi\iu/3)$ a third root of unity, that controls the evolution
\eq{
\begin{pmatrix}
    \ha_1^\dagger\\
    \ha_2^\dagger\\
    \ha_3^\dagger
\end{pmatrix}
\to
\mathbf{M}^\top 
\begin{pmatrix}
    \ha_1^\dagger\\
    \ha_2^\dagger\\
    \ha_8^\dagger\\
    \ha_{14}^\dagger\\
    \ha_{20}^\dagger
\end{pmatrix}.
} \sugg{The top three highlighted rails on the left of Fig.~\ref{fig:circuit diagram unravel} that begin without the vacuum state are thus transferred into the five highlighted output rails on the right of Fig.~\ref{fig:circuit diagram unravel}; the other two highlighted input rails house vacuum states.}
The components of the full verified transfer matrix are recorded in Appendix~\ref{sec:validating transfer matrix}. Such a transfer matrix will amplify \textit{any} state input in mode $3$ and teleport it to mode 2, should the heralding events be successful.

We verify the action of this circuit on our chosen input states in Appendix~\ref{sec:validating transfer matrix}, although the result is guaranteed by Ref. \cite{Guanzonetal2022}. Conditional on one of the successful heralding patterns on the applicable four modes, the input state should be teleamplified to one of
\eq{
\ket{g\psi;2}\propto \ket{0}-\tanh r \sqrt{2}\ket{2}\frac{1-\tau}{\tau}\omega^l,
\label{eq:amp state from TMSV with phase}
} where the phase is determined by the heralding pattern $\ket{0,1,1}$ ($l=0$), $\ket{1,0,1}$ ($l=1$), or $\ket{1,1,0}$ ($l=2$) in the outputs of the 8th, 14th, and 20th time bins. \sugg{Just as in a standard quantum teleportation scenario, the classical information ``$l$'' must be conveyed to the party receiving the teleamplified state, who can undo the extra effect of this phase $\omega^l$ with a deterministic operation that depends on $l$. In our work we do not independently verify this phase relationship and can at best rely on Borealis previously reporting phase deviations of less than $0.03$ radians in the first two delay loops for an overall phase noise of less than one part per 100 billion \cite{Madsenetal2022}. The absence of feedforwarding on this device makes it useful for inspecting the teleamplification algorithm up to the final rephasing; alternatively, one could ignore all postselection events other than the pattern corresponding to $l=0$, at the cost of a reduced probability of successful heralding.}

 The gain parameter $g=\sqrt{\frac{1-\tau}{\tau}}$ is set by one programmable transmissivity that we vary between 1/8 and 128. We run the circuit multiple times to build sufficient statistics among the successful heralding events, whose success probability depends on the squeezing parameter $r$ and the gain $g$. We run $4\times 10^6$ trials for each of the gain parameters $g=1/2$ (attenuation), $g=1$ (no gain), $g=2$, and $g=4$ with squeeze parameter $r=1.148$. Other squeeze and gain parameters were also run but fewer numbers of times, with the measured data all available via Github (see Appendix~\ref{sec:details} for a copy of the device certificate for $r=1.148$ and other setup details).

\section{Results}
We use the photon-number distribution of the teleported mode to evaluate the success of the amplification.\sugg{\footnote{\sugg{Borealis only provides access to photon-number distributions, so full tomography is out of the question. One could imagine obtaining relative-phase information between the coefficients $\psi_k$ by interfering the teleamplified state with another beam of light. Since Borealis only makes one non-vacuum state and only makes those states in adjacent modes, that beam would have to be a SMSV in the fourth rail of Fig.~\ref{fig:circuit diagram unravel} or a set of SMSVs in the fourth through $m$th rails. But to interfere with the outgoing second mode after it has passed through the entire circuit, the second mode must be delayed at the third delay loop and wait to interfere with the $m=$38th mode. Adding SMSV states in all of the intervening modes ruins the teleamplification setup, so full state tomography is precluded.}}} Since this does not depend on the heralding pattern [i.e., it is independent from the integer $l$ in Eq.~\eqref{eq:amp state from TMSV with phase}], we aggregate results from all three heralding patterns. \sugg{In this sense we are only verifying the amplification part of the scheme and not its phase-insensitivity because the latter is not directly measurable with this setup.} In total, we measure 23084, 12773, 4555, and 2459 successful postselection events for the gain parameters $g=1/2$, $1$, $2$, and $4$, respectively \sugg{(again, all of these data are available on Github)}. The heralded states from those events are what we now characterize.

\begin{figure}
    \centering
    \includegraphics[width=\columnwidth]{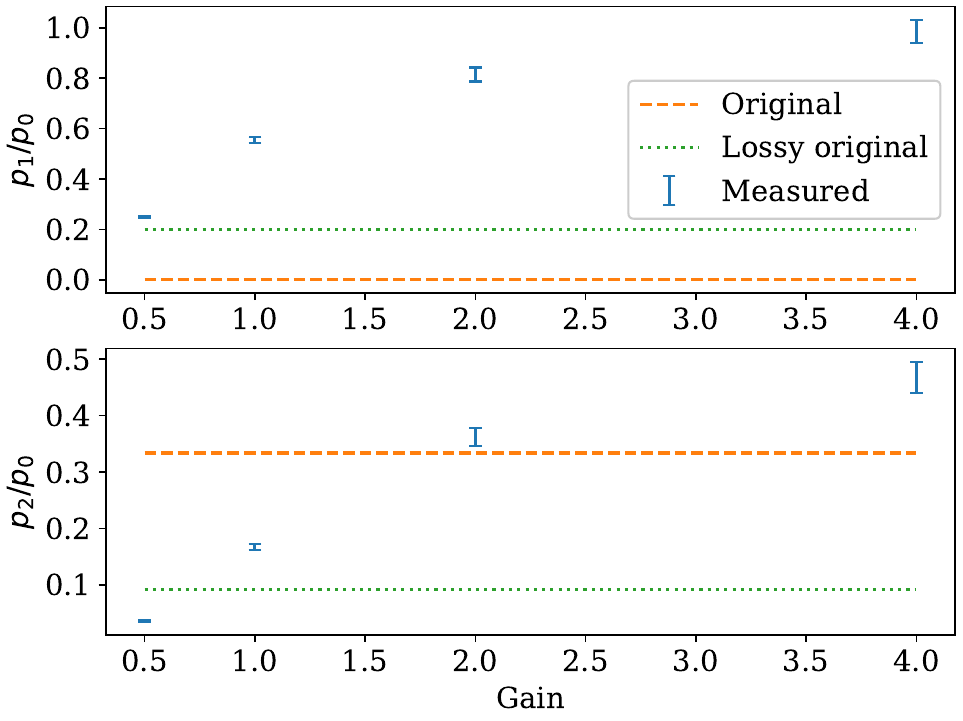}
    \caption{Measured ratios of the photon-number probabilities for the teleamplified state versus gain. The one-photon probability is expected to scale quadratically with gain and the two-photon probability quartically, but they each grow slower than linearly. For comparison, we plot the original ratios of these probabilities, showing that indeed there is gain that increases for increasing gain parameter, relative to either the perfect initial state or the true, lossy initial state. \sugg{While we have a model for these values as a function of the entire setup, future work could seek a simplified equation to describe this gain as a function of experimental imperfections.}
    The error bars are given by Poisson statistics on the number of expected counts. \sugg{Quantities in this and all plots are dimensionless; here ``gain'' is the factor $g$ in an ideal amplification process $\ket{\alpha}\to\ket{g\alpha}$.}}
    \label{fig:gains}
\end{figure}

We observe the photon-number distributions of the teleamplified states to increase their single- and two-photon components relative to their zero-photon component as the gain increases (Fig.~\ref{fig:gains}\sugg{; the green dotted line will be explained later}). These demonstrate successful teleamplification. \sugg{For the purpose of uncertainty calculations, we take the detector readouts with complete confidence, which is a good approximation because their intrinsic dark counts are negligible at less than $10^{-10}$ per detection event \cite{Milleretal2003} and they can resolve between 0 and 2 photons (for all of the heralding patterns) with greater than 0.99999999 probability of success and up to 7 photons (the highest depicted in any plot here) with probability greater than 0.999  \cite{Morais2022arxiv}. The dominant errors by far then come from loss. For example, the minimum detector efficiency is 0.893; since these parameter are known from device calibration (again, see Github for device certificates), they do not contribute to the error bars depicted in Fig.~\ref{fig:gains}. Rather, we assume errors on our photon-number counts arise from Poissonian photodetection statistics and take the rest of the circuit to behave as anticipated including pervasive loss.}

One might have expected there to be no single-photon contribution present \sugg{in the amplified state as in Fig.~\ref{fig:gains}}, but loss in the initial state prior to being amplified gives rise to an initial nonzero single-photon contribution that can then be amplified. Loss that occurs later in the circuit degrades the fidelity between the heralding pattern and the true desired postselection events, which also gives rise to the teleamplified state having small nonzero contributions from photon numbers greater than $n=2$.

\begin{figure}
    \centering
    \includegraphics[width=\columnwidth]{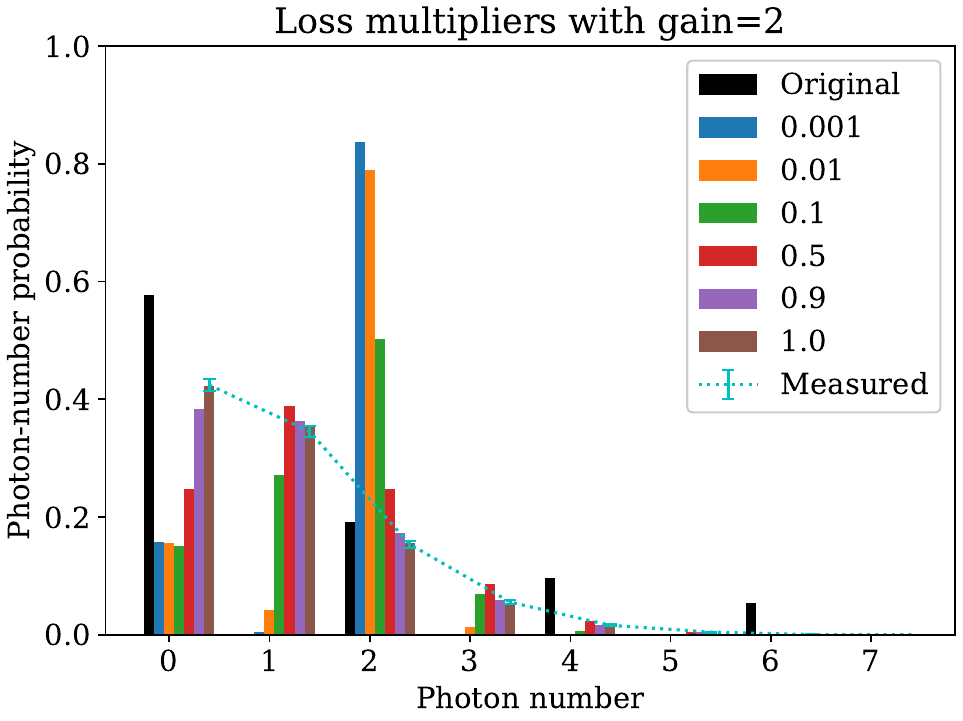}
    \caption{Measured photon-number distribution for the amplified state in the teleported mode (cyan, dashed line; Poissonian error bars) compared to the initial distribution in the input mode (far left bar in each cluster, black). Also plotted are simulated results for the teleamplified state, should all of the loss parameters present in the circuit have been multiplied by a factor between 0.001 (blue, second bar from left) increasing to the full loss present 1.0 (brown, far right bar). Here the gain parameter is $g=2$ and the data match the predictions, with the two-photon component increasing relative to the zero-photon component. {The fidelity between the measured and ideal (lossless) photon-number distributions is $0.83\pm 0.01$ and was expected to be $0.84$.}}
    \label{fig:g2 perfect orig}
\end{figure}

How does the measured gain compare to the expected value? Initially, one expects a quadratic increase with gain for the relative single-photon probability and a quartic increase for the two-photon probability. However, loss in the heralding modes prior to their detection causes mismatches in the expected distributions: when a detector registers ``one photon,'' for example, it is possible that there were actually two photons in that mode and one got lost along the way\sugg{, either due to mode mismatch or detector inefficiency or other sources of loss}. We can use calibration data from all of the components of Borealis to plot the true expected photon-number distributions from the lossy circuit. The case for $g=2$ is plotted in Fig.~\ref{fig:g2 perfect orig} with the rest of the data deferred to Appendix~\ref{sec:perfect figs}. We observe excellent agreement between the predicted and measured values for all of the photon numbers in the teleamplified state.

\subsection{\sugg{Effects of loss}}
\sugg{Inspecting Fig.~\ref{fig:circuit diagram unravel}, there are multiple places for loss to occur: in the delay loop for the nearest-neighbour beam splitters, the delay loop for the 6th-nearest-neighbour beam splitter, the rest of the free propagation, and detector inefficiencies that can be modelled as loss.}
We can simulate the same setup with smaller loss parameters to see how the photon-number distribution in the output mode is affected by loss. We collectively reduce all of the loss parameters throughout the circuit by a factor of $q\in (0, 1)$, which is like increasing the relevant transmission parameters as $\eta\to\eta+(1-\eta)(1-q)$. Expected photon-number distributions for different values of $q$ are also plotted in Fig.~\ref{fig:g2 perfect orig}, where we can compare the transition between tiny loss probabilities ($q=0.001$) and the true loss probabilities ($q=1$). As the amount of loss increases, the zero- and one-photon components of the state increase, requiring \textit{tiny} loss parameters $q\lesssim 0.001$ for the one-photon component to vanish. For the two-photon component to reach as large of a probability as the ideal teleamplified value, small loss parameters $q\lesssim 0.01$ are also required. These are due to the sensitivity of such heralded schemes to loss, as such small loss parameters would require each component of Borealis to have transmission parameters and efficiencies greater than $0.999$.

\sugg{Again inspecting Fig.~\ref{fig:circuit diagram unravel}, we see that the state-to-be-amplified $\ket{\psi}$ must traverse the first delay loop 11 times before interacting with another vacuum mode, by virtue of Borealis requiring the initially populated modes to be adjacent time bins. By taking the calibrated transmission probability for this loop from Borealis, 0.88, we can model the input state instead as the one modified from $\ket{\psi}$ by first transforming $\ha\to \sqrt{0.88^{11}}\ha+\sqrt{1-0.88^{11}}\hb$ and then tracing out the mode annihilated by $\hb$ that began in the vacuum state. This is how we obtain the green dotted lines in Fig.~\ref{fig:gains} with nonzero initial probability of finding a single photon; we refer to the ``lossy original'' state that is effectively an attenuated SMSV. }

\begin{figure}
    \centering
    \includegraphics[width=\columnwidth]{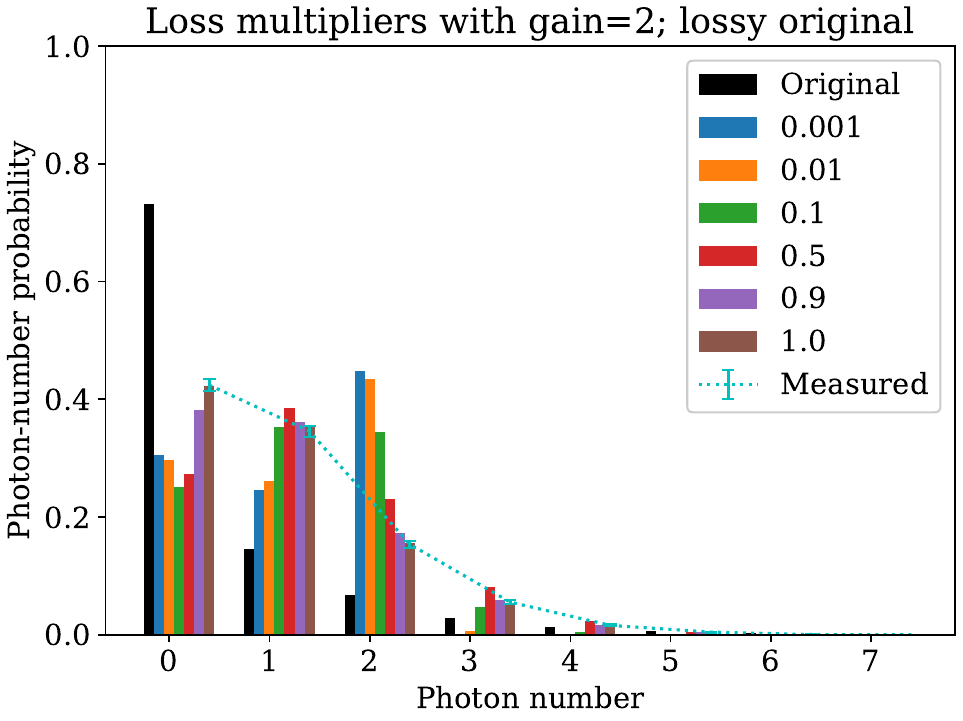}
    \caption{Same plot as Fig.~\ref{fig:g2 perfect orig}: photon-number distribution for the amplified state in the teleported mode (cyan, dashed line; Poissonian error bars) compared to the initial distribution in the input mode (far left bar in each cluster, black), but now considering the input state to have been the attenuated squeezed state that resulted from loss in the first delay line before interacting with the rest of the circuit. The simulated results for the teleamplified state thus change relative to Fig.~\ref{fig:g2 perfect orig} and are much closer to the measured values. Regardless, when the loss parameters are set to be the same as in the actual circuit, the data match the predictions (the far right bars in the histograms are the same as in Fig.~\ref{fig:g2 perfect orig}). Here the gain parameter is $g=2$ and the data match the predictions, where now it is more clear how the one- and two-photon components increase in the amplified state relative to the input state.}
    \label{fig:g2 lossy orig}
\end{figure}

Another reason for the sensitivities of this setup is the requirement to repeatedly store light in delay loops so that it can interact with the appropriate mode. We can use this to our advantage by again considering the initial state not to be the perfect SMSV but the lossy version that results from being stored in the first delay loop 11 times.\footnote{We remark that a SMSV subject to photon lossy is still nonclassical according to its $P$ function, although nonclassicality of our input states is merely a curiosity and is by no means necessary.} This initial state is plotted in Fig.~\ref{fig:g2 lossy orig} with the measured data for $g=2$, where it is now clear that the one-photon and two-photon components are amplified and the zero-photon component is diminished relative to the original state.
We can then take the lossy initial state and see what would come of the setup should the rest of the components of the circuit have their loss probabilities diminish by a factor $q$; these are also plotted in Fig.~\ref{fig:g2 lossy orig} \sugg{and are equivalent to the simulations performed to attain Fig.~\ref{fig:g2 perfect orig} but with the loss parameters undiminished for the state $\ket{\psi}$ that is initially held in the first delay loop}. Now it is seen that small loss parameters with $q\lesssim 0.01$ are required to observe the true amplified distribution, which is less stringent than $0.001$ from before but still shows how sensitive this heralded scheme is to photon loss. 

\begin{figure}
    \centering
    \includegraphics[width=\columnwidth]{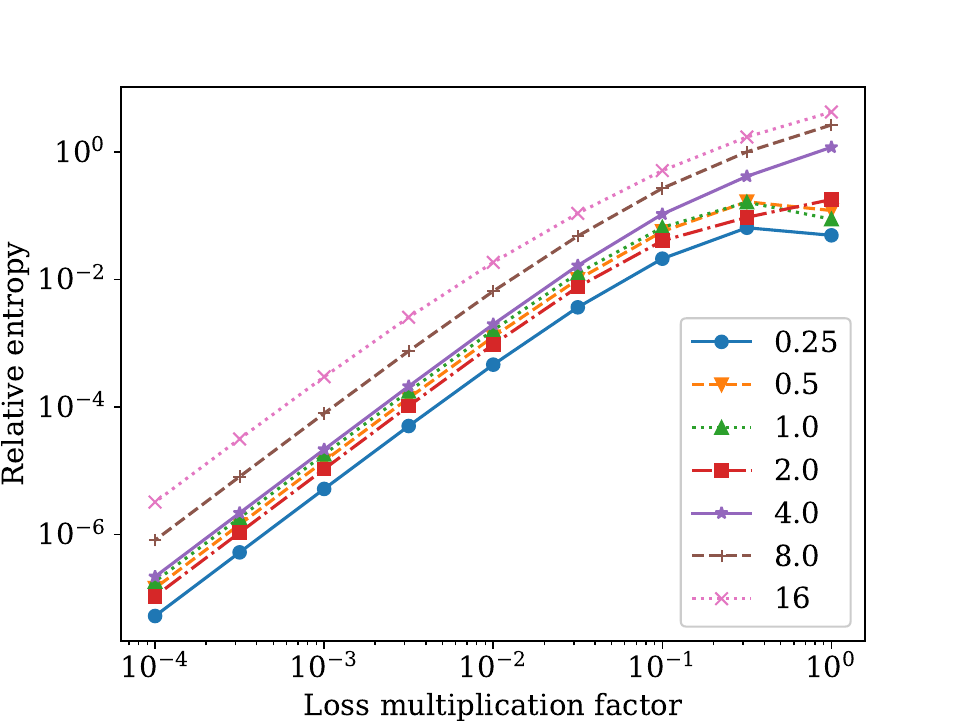}
    \caption{Similarity between the ideal amplified photon-number distribution and the predicted teleamplified photon-number distribution when the loss in the circuit is decreased throughout by a multiplicative factor. The relative entropies approach their minimum value of 0 quadratically with this factor, as expected from a minimum, by the time the losses are multiplied by $\sim 0.01 - 0.1$. Lower gain parameters are more loss tolerant, with monotonic behaviour other than the one exception that $g=2$ is more loss tolerant than $g=1$ and $g=1/2$ once the loss is decreased by a factor of 10.}
    \label{fig:KL}
\end{figure}

\begin{figure}
    \centering
    \includegraphics[width=\columnwidth]{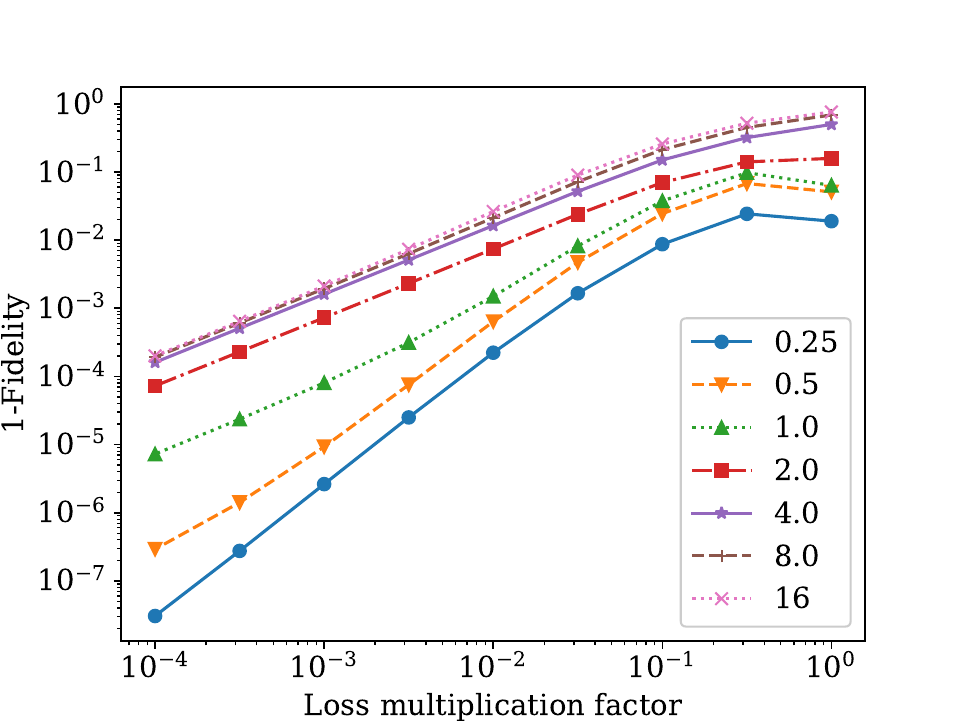}
    \caption{Same as Fig.~\ref{fig:KL} but plotting \sugg{photon-number probability distributions'} fidelity instead of relative entropy. The fidelities approach their maximum value of 1 quadratically with this factor, as expected from a maximum, by the time the losses are multiplied by $\sim 0.01 - 0.1$. Lower gain parameters are more loss tolerant, with monotonic behaviour rectifying the exceptions from Fig.~\ref{fig:KL}.}
    \label{fig:fidelity}
\end{figure}

We conclude by looking at the same simulated photon-number distributions versus loss for the other gain parameters and the lossy initial state. It roughly seems that the larger gain parameters are less loss tolerant when we look at the same figures, all recorded in Appendix~\ref{sec:lossy figs}, but such a qualitative reasoning is perilous and so the figures are less informative. To quantify this loss tolerance, we compute the relative entropy (i.e., the Kullback-Leibler divergence) between the simulated photon-number distributions' zero-, one-, and two-photon components with various amounts of loss to the distribution without loss, all for the lossy initial state. In Fig.~\ref{fig:KL}, we plot these divergences, which go to zero approximately quadratically with the loss multiplication factor. It is there evident that larger gain parameters are less loss tolerant, with smaller gain parameters generally achieving lower relative entropies for the same value of loss. We also plot the fidelities expected for different amounts of loss in Fig.~\ref{fig:fidelity}, confirming that larger gain parameters are more sensitive to loss. \sugg{These fidelities are calculated between photon-number distributions and, since the simulations maintain phase coherence, are equivalent to the best quantum fidelities that could be achieved for the various gain and loss parameters. We thus conclude that an interferometer maintaining phase coherence could achieve a gain of $g=16$ with fidelity greater than 99$\%$ if the overall losses are all multiplied by 0.01, corresponding to detector efficiencies greater than 0.999 and common-mode transmission probabilities greater than 0.99.}

\section{Conclusions}
We have demonstrated amplification of an attenuated squeezed state up to its $n=2$ cutoff and teleportation of the state in question by adapting Ref. \cite{Guanzonetal2022}'s heralded scheme to the programmable quantum device Borealis, for various gain parameters, with one- and two-photon components of the state growing by a factor of up to ~5 in probability. Our fidelities with the ideal (lossless) states for gain parameters of $1/2$, $1$, $2$, and $4$ were $93.2(6)\%$, $92.6(9)\%$, $83(1)\%$, and $50(1)\%$, respectively. We also presented simulations showing how sensitive this heralded scheme is to loss in the heralding modes; a device dedicated only to teleamplification could circumvent some loss by avoiding delay loops, but ultimately would be just as sensitive to overall loss as our results. Relative to the actual predicted values in the presence of loss, our fidelities are always greater than 98$\%$. \sugg{The outlook is that, once teleamplification's interferometer and PNRD components are created, the main limiting factor for the protocol is photon loss. Improvements in the form of enhanced mode matching or heightened detector efficiencies will help make this protocol viable for varied applications.} Teleamplification, an ideal form of noiseless linear amplification, has many applications across quantum information science \cite{Gisinetal2010,Ralph2011,Blandinoetal2012,KellererRibak2016,Zhaoetal2017,Diasetal2020,Zhouetal2020}, showcasing the usefulness and tunability of the machines and components developed for fault tolerant quantum computation to a wide variety of problems and solutions.

\begin{acknowledgments}
The authors acknowledge that the NRC headquarters is located on the traditional unceded territory of the Algonquin Anishinaabe and Mohawk people, as well as support from NRC's Quantum Sensors Challenge Program. {This research was made possible through Innovative Solutions Canada’s Testing Stream by providing
access to Xanadu’s Borealis device.} AZG acknowledges funding from the NSERC PDF program. {KH acknowledges funding from NSERC's Discovery Grant.}
\end{acknowledgments}

\clearpage
\onecolumngrid

\appendix
\section{\sugg{Simplified circuit diagram}}
\label{sec:simplified circuit diagram}
\begin{figure*}
    \centering
    \includegraphics[width=\textwidth,trim={0 9.5cm 0 0},clip]{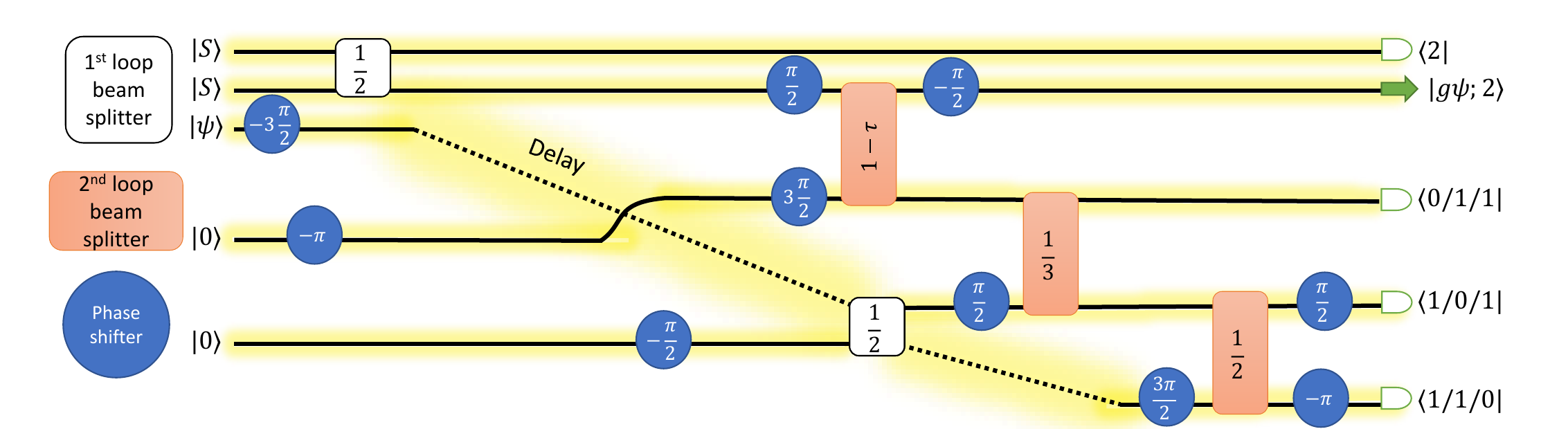}
    \caption{\sugg{Circuit programmed for Borealis showing only the five crucial modes from Fig.~\ref{fig:circuit diagram unravel}. Delays are depicted by dashed lines, which are physically enacted using reflective beam splitters that prevent the light from escaping the delay loop. }  }
    \label{fig:circuit condensed}
\end{figure*}

\sugg{A simplified version of Fig.~\ref{fig:circuit diagram unravel} that ignores the unhighlighted components is depicted in Fig.~\ref{fig:circuit condensed}. There are five crucial time bin modes unravelled from top to bottom for visualization purposes. }

\section{Validating the overall transfer matrix}
\label{sec:validating transfer matrix}
The circuit that we ran had the following verified transfer matrix acting on all 20 time-bin modes:
%\begin{widetext}
    \begin{equation}
    \mathbf{U}=\left(
\begin{array}{cccccccccccccccccccc}
 \frac{1}{\sqrt{2}} & \frac{i}{\sqrt{2}} & 0 & 0 & 0 & 0 & 0 & 0 & 0 & 0 & 0 & 0 & 0 & 0 & 0 & 0 & 0 & 0 & 0 & 0 \\
 \frac{i \sqrt{1-\tau}}{\sqrt{2}} & \frac{\sqrt{1-\tau}}{\sqrt{2}} & 0 & 0 & 0 & 0 & 0 & 0 & -\sqrt{\tau} & 0 & 0 & 0 & 0 & 0 & 0 & 0 & 0 & 0 & 0 & 0 \\
 0 & 0 & 0 & 1 & 0 & 0 & 0 & 0 & 0 & 0 & 0 & 0 & 0 & 0 & 0 & 0 & 0 & 0 & 0 & 0 \\
 0 & 0 & 0 & 0 & 1 & 0 & 0 & 0 & 0 & 0 & 0 & 0 & 0 & 0 & 0 & 0 & 0 & 0 & 0 & 0 \\
 0 & 0 & 0 & 0 & 0 & 1 & 0 & 0 & 0 & 0 & 0 & 0 & 0 & 0 & 0 & 0 & 0 & 0 & 0 & 0 \\
 0 & 0 & 0 & 0 & 0 & 0 & 1 & 0 & 0 & 0 & 0 & 0 & 0 & 0 & 0 & 0 & 0 & 0 & 0 & 0 \\
 0 & 0 & 0 & 0 & 0 & 0 & 0 & 1 & 0 & 0 & 0 & 0 & 0 & 0 & 0 & 0 & 0 & 0 & 0 & 0 \\
 -\frac{i \sqrt{\tau}}{\sqrt{6}} & -\frac{\sqrt{\tau}}{\sqrt{6}} & {-}\frac{1}{\sqrt{3}} & 0 & 0 & 0 & 0 & 0 & -\frac{\sqrt{1-\tau}}{\sqrt{3}} & 0 & 0 & 0 & 0 & 0 & {-}\frac{1}{\sqrt{3}} & 0 & 0 & 0 & 0 & 0 \\
 0 & 0 & 0 & 0 & 0 & 0 & 0 & 0 & 0 & 1 & 0 & 0 & 0 & 0 & 0 & 0 & 0 & 0 & 0 & 0 \\
 0 & 0 & 0 & 0 & 0 & 0 & 0 & 0 & 0 & 0 & 1 & 0 & 0 & 0 & 0 & 0 & 0 & 0 & 0 & 0 \\
 0 & 0 & 0 & 0 & 0 & 0 & 0 & 0 & 0 & 0 & 0 & 1 & 0 & 0 & 0 & 0 & 0 & 0 & 0 & 0 \\
 0 & 0 & 0 & 0 & 0 & 0 & 0 & 0 & 0 & 0 & 0 & 0 & 1 & 0 & 0 & 0 & 0 & 0 & 0 & 0 \\
 0 & 0 & 0 & 0 & 0 & 0 & 0 & 0 & 0 & 0 & 0 & 0 & 0 & 1 & 0 & 0 & 0 & 0 & 0 & 0 \\
 \frac{i \sqrt{\tau}}{\sqrt{6}} & \frac{\sqrt{\tau}}{\sqrt{6}} & {-}\frac{1}{6} \left(3\iu+\sqrt{3}\right) & 0 & 0 & 0 & 0 & 0 & \frac{\sqrt{1-\tau}}{\sqrt{3}} & 0 & 0 & 0 & 0 & 0 & {-}\frac{1}{6} \left(-3\iu+\sqrt{3}\right) & 0 & 0 & 0 & 0 & 0 \\
 0 & 0 & 0 & 0 & 0 & 0 & 0 & 0 & 0 & 0 & 0 & 0 & 0 & 0 & 0 & 1 & 0 & 0 & 0 & 0 \\
 0 & 0 & 0 & 0 & 0 & 0 & 0 & 0 & 0 & 0 & 0 & 0 & 0 & 0 & 0 & 0 & 1 & 0 & 0 & 0 \\
 0 & 0 & 0 & 0 & 0 & 0 & 0 & 0 & 0 & 0 & 0 & 0 & 0 & 0 & 0 & 0 & 0 & 1 & 0 & 0 \\
 0 & 0 & 0 & 0 & 0 & 0 & 0 & 0 & 0 & 0 & 0 & 0 & 0 & 0 & 0 & 0 & 0 & 0 & 1 & 0 \\
 0 & 0 & 0 & 0 & 0 & 0 & 0 & 0 & 0 & 0 & 0 & 0 & 0 & 0 & 0 & 0 & 0 & 0 & 0 & 1 \\
 -\frac{i \sqrt{\tau}}{\sqrt{6}} & -\frac{\sqrt{\tau}}{\sqrt{6}} & {-}\frac{1}{6} \left(3\iu-\sqrt{3}\right) & 0 & 0 & 0 & 0 & 0 & -\frac{\sqrt{1-\tau}}{\sqrt{3}} & 0 & 0 & 0 & 0 & 0 & {-}\frac{1}{6} \left(-3\iu-\sqrt{3}\right) & 0 & 0 & 0 & 0 & 0 \\
\end{array}
\right).
\end{equation}
%\end{widetext} 
The phases in each row are actually irrelevant to the output photon-number distributions, as are the phases for all of the input vacuum states and changes in phase by $\pi$ on the input SMSV states. The relevant modes with entries other than unity correspond to the highlighted wires in Fig.~\ref{fig:circuit diagram unravel}. \sugg{Many of the phases in Fig.~\ref{fig:circuit diagram unravel} do not affect the dynamics of our setup but are chosen so that many of the entries of $\mathbf{U}$ are equal to 1.}

We can verify the action of this transfer matrix on our input states, although one can prove in general that it will teleport and amplify any input state should the proper heralding pattern be observed. The initial state has an SMSV in each of the first three modes \eq{
\ket{\Psi}\propto\exp[-\tanh r \eu^{\iu\varphi}  (\ha^{2\dagger}_1+\ha^{2\dagger}_2+\ha^{2\dagger}_3)]\vac.
} This transforms into
\eq{
\ket{\Psi^\prime}\propto&\exp\{-\tanh r \eu^{\iu\varphi} [(M_{11}\had_1+M_{21}\had_2+M_{81}\had_8+M_{14,1}\had_{14}+M_{20,1}\had_{20})^2
\\
&
+(M_{12}\had_1+M_{22}\had_2+M_{82}\had_8+M_{14,2}\had_{14}+M_{20,2}\had_{20})^2
+(M_{13}\had_1+M_{23}\had_2+M_{83}\had_8+M_{14,3}\had_{14}+M_{20,3}\had_{20})^2]\}\vac\\
=&\exp\{-\tanh r \eu^{\iu\varphi} [(M_{11}\had_1+M_{21}\had_2+M_{81}\had_8+M_{14,1}\had_{14}+M_{20,1}\had_{20})^2\\
&
+(\iu M_{11}\had_1-\iu M_{21}\had_2-\iu M_{81}\had_8-\iu M_{14,1}\had_{14}-\iu M_{20,1}\had_{20})^2
+(M_{83}\had_8+M_{14,3}\had_{14}+M_{20,3}\had_{20})^2]\}\vac\\
=&\exp\{-\tanh r \eu^{\iu\varphi} [4M_{11}\had_1(M_{21}\had_2+M_{81}\had_8+M_{14,1}\had_{14}+M_{20,1}\had_{20})]\}\\
&\times\exp \{-\tanh r \eu^{\iu\varphi}[(M_{83}\had_8+M_{14,3}\had_{14}+M_{20,3}\had_{20})^2]\}\vac.
} Projecting the first mode onto the $2$-photon state is straightforward because $\had_1$ appears only once:
\eq{
\bra{2}_1\ket{\Psi^\prime}\propto &(M_{21}\had_2+M_{81}\had_8+M_{14,1}\had_{14}+M_{20,1}\had_{20})^2\exp \{-\tanh r \eu^{\iu\varphi}[(M_{83}\had_8+M_{14,3}\had_{14}+M_{20,3}\had_{20})^2]\}\vac\\
&=\ket{0}_2(M_{81}\had_8+M_{14,1}\had_{14}+M_{20,1}\had_{20})^2\exp \{-\tanh r \eu^{\iu\varphi}[(M_{83}\had_8+M_{14,3}\had_{14}+M_{20,3}\had_{20})^2]\}\vac+\\
&2M_{21}\ket{1}_2(M_{81}\had_8+M_{14,1}\had_{14}+M_{20,1}\had_{20})\exp \{-\tanh r \eu^{\iu\varphi}[(M_{83}\had_8+M_{14,3}\had_{14}+M_{20,3}\had_{20})^2]\}\vac+\\
&M_{21}^2\sqrt{2}\ket{2}_2\exp \{-\tanh r \eu^{\iu\varphi}[(M_{83}\had_8+M_{14,3}\had_{14}+M_{20,3}\had_{20})^2]\}\vac.
} Modes $8$, $14$, and $20$ get projected onto having a total of two photons, so overall we only need to consider the terms
\eq{\ket{0}_2(M_{81}\had_8+M_{14,1}\had_{14}+M_{20,1}\had_{20})^2\vac+M_{20}^2\sqrt{2}\ket{2}_2\{-\tanh r \eu^{\iu\varphi}[(M_{83}\had_8+M_{14,3}\had_{14}+M_{20,3}\had_{20})^2]\}\vac.
} We find the three patterns
\eq{
\bra{2}_1\otimes\bra{0}_8\otimes\bra{1}_{14}\otimes\bra{1}_{20}\ket{\Psi^\prime}&\propto 2M_{14,1}M_{20,1}\ket{0}_2 -\tanh r \eu^{\iu\varphi}2M_{14,3}M_{20,3}M_{20}^2\sqrt{2}\ket{2}_2\\
&\propto  \ket{0}_2 -\tanh r \eu^{\iu\varphi}\sqrt{2}(\frac{1-\tau}{\tau})\ket{2}_2,
}
\eq{
\bra{2}_1\otimes\bra{1}_8\otimes\bra{0}_{14}\otimes\bra{1}_{20}\ket{\Psi^\prime}&\propto 2M_{81} M_{20,1}\ket{0}_2 -\tanh r \eu^{\iu\varphi}2M_{83}M_{20,3}M_{20}^2\sqrt{2}\ket{2}_2\\
&\propto 2(-\frac{\tau}{6})\ket{0}_2 -\tanh r \eu^{\iu\varphi}2\frac{\eu^{2\iu\pi/3}}{3}(-\frac{1-\tau}{2})\sqrt{2}\ket{2}_2\\
&\propto \ket{0}_2 -\tanh r \eu^{\iu\varphi}\sqrt{2}(\eu^{2\iu\pi/3}\frac{1-\tau}{\tau})\ket{2}_2,
}
and
\eq{
\bra{2}_1\otimes\bra{1}_8\otimes\bra{1}_{14}\otimes\bra{0}_{20}\ket{\Psi^\prime}&\propto \ket{0}_2 2M_{81}M_{14,1}\vac+M_{20}^2\sqrt{2}\ket{2}_2\{-\tanh r \eu^{\iu\varphi}2M_{83}M_{14,3}\}\vac\\
&\propto \ket{0}_2 2\frac{\tau}{6}+(-\frac{1-\tau}{2})\sqrt{2}\ket{2}_2\{-\tanh r \eu^{\iu\varphi}2\frac{\eu^{\iu\pi/3}}{3}\}\\
&\propto \ket{0}_2 -\tanh r \eu^{\iu\varphi}\sqrt{2}(\eu^{-2\iu\pi/3}\frac{1-\tau}{\tau})\ket{2}_2.
} All three correspond to a gain relative to the squeezed state $\ket{0}-\tanh r \eu^{\iu\varphi}\sqrt{2}\ket{2}$ by a factor $g=\sqrt{\frac{1-\tau}{\tau}}$, with different phases for the different heralded patterns that can be adjusted with classical communication as in standard teleportation schemes. The actual initial state had $\varphi=0$ but, as iterated above, this circuit works to teleport and amplify any input state.

\section{Device details}
\label{sec:details}
A full diagram with full details of the device used to perform our demonstration can be found in Ref.~\cite{Madsenetal2022}. The certificate from the device on the day the experiment was run was \texttt{\{'finished\_at': '2023-05-29T17:41:34.623450+00:00', 'target': 'borealis', 'loop\_phases': [1.268, -0.051, 1.848], 'schmidt\_number': 1.151, 'common\_efficiency': 0.386, 'loop\_efficiencies': [0.88, 0.879, 0.793], 'squeezing\_parameters\_mean': \{'low': 0.678, 'high': 1.148, 'medium': 1.06\}, 'relative\_channel\_efficiencies': [0.918, 0.938, 0.912, 1.0, 0.961, 0.917, 0.893, 0.969, 0.951, 0.955, 0.965, 0.998, 0.947, 0.966, 0.947, 0.898]\}}.

\section{Figures for other gain parameters}
\label{sec:perfect figs}
We here list the same figures as in the main text but for gain parameters of $g=1/2$, $1$ and $4$. These include the measured data for each parameter as well as the predicted photon-number distributions when reducing the loss by various multiplicative factors.
\subsection{Squeezed state input}
\begin{figure}
    \centering
    \includegraphics[width=0.5\columnwidth]{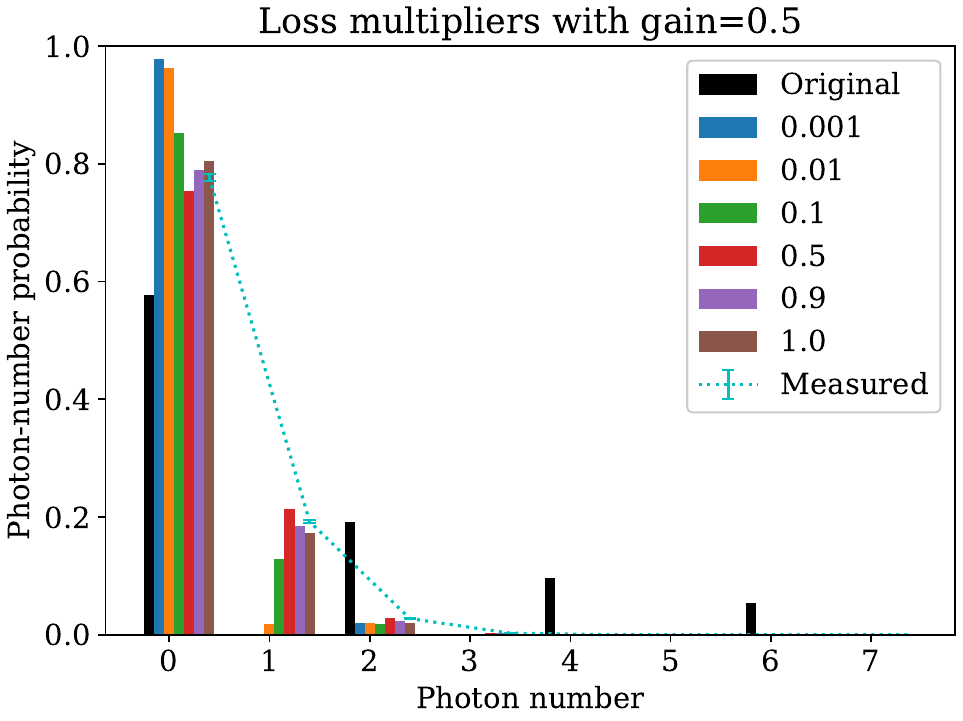}
    \caption{Same as Fig.~\ref{fig:g2 perfect orig} but with gain parameter $g=1/2$. {The fidelity between the measured and ideal (lossless) photon-number distributions is $0.932\pm 0.006$ and was expected to be $0.949$.}}
    \label{fig:g05 perfect orig}
\end{figure}

\begin{figure}
    \centering
    \includegraphics[width=0.5\columnwidth]{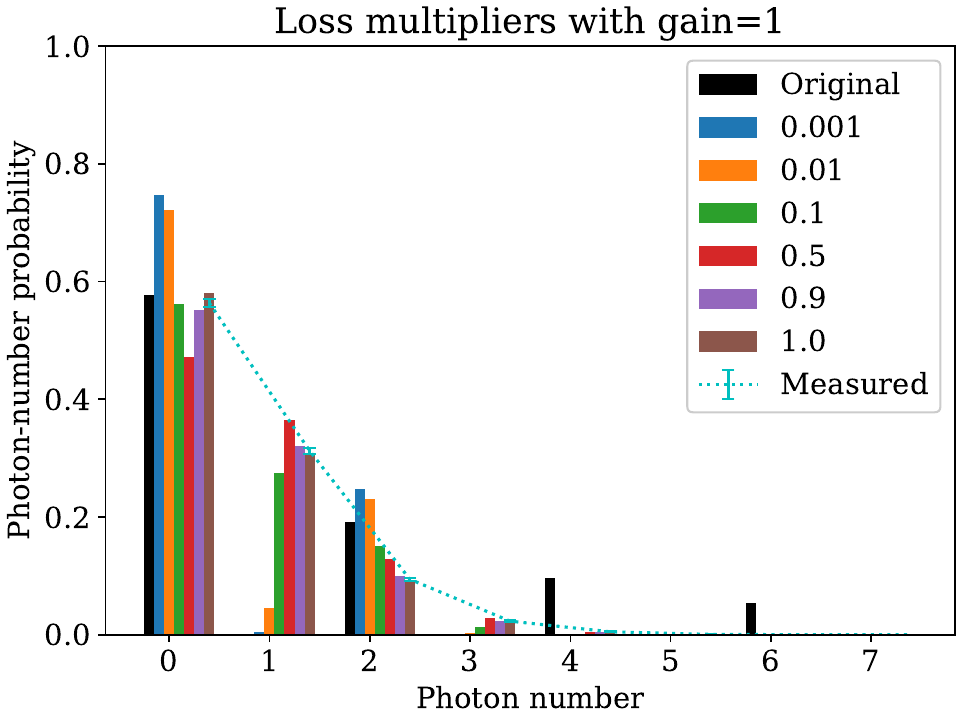}
    \caption{Same as Fig.~\ref{fig:g2 perfect orig} but with gain parameter $g=1$. {The fidelity between the measured and ideal (lossless) photon-number distributions is $0.926\pm0.009$ and was expected to be $0.936$.}}
    \label{fig:g1 perfect orig}
\end{figure}

\begin{figure}
    \centering
    \includegraphics[width=0.5\columnwidth]{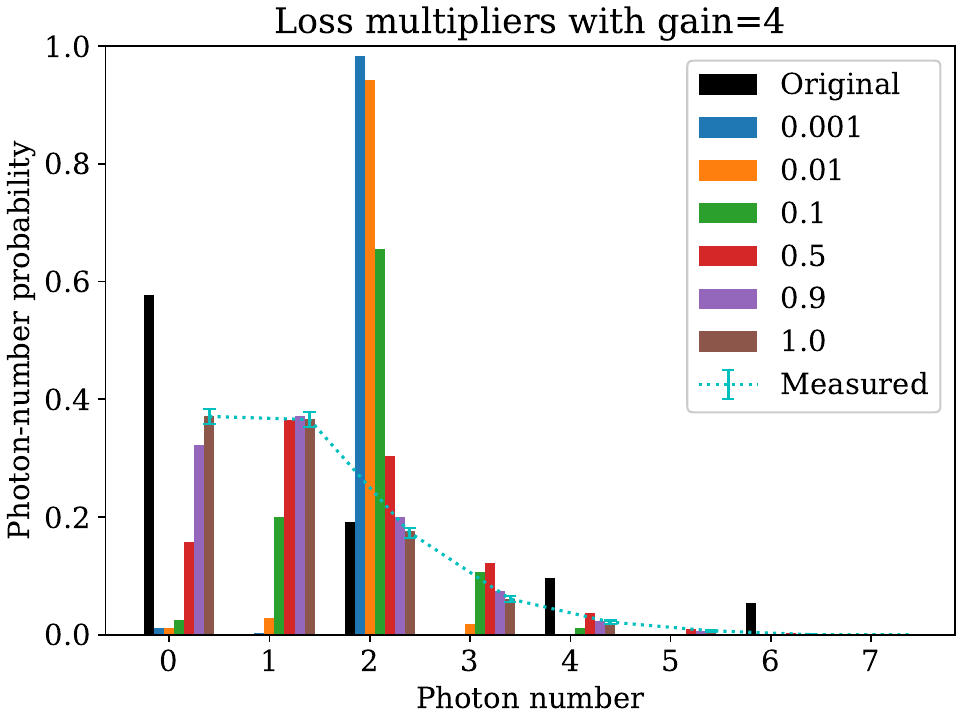}
    \caption{Same as Fig.~\ref{fig:g2 perfect orig} but with gain parameter $g=4$. {The fidelity between the measured and ideal (lossless) photon-number distributions is $0.50\pm 0.01$ and was expected to be $0.50$.}}
    \label{fig:g4 perfect orig}
\end{figure}

We begin by reducing the loss by multiplicative factors, considering the original state to be a perfect SMSV. The figures for gains of $g=1/2$, $1$ and $4$ are plotted in Figs.~\ref{fig:g05 perfect orig},~\ref{fig:g1 perfect orig}, and \ref{fig:g4 perfect orig}, respectively.

\subsection{Attenuated squeezed state input}
\label{sec:lossy figs}
\begin{figure}
    \centering
    \includegraphics[width=0.5\columnwidth]{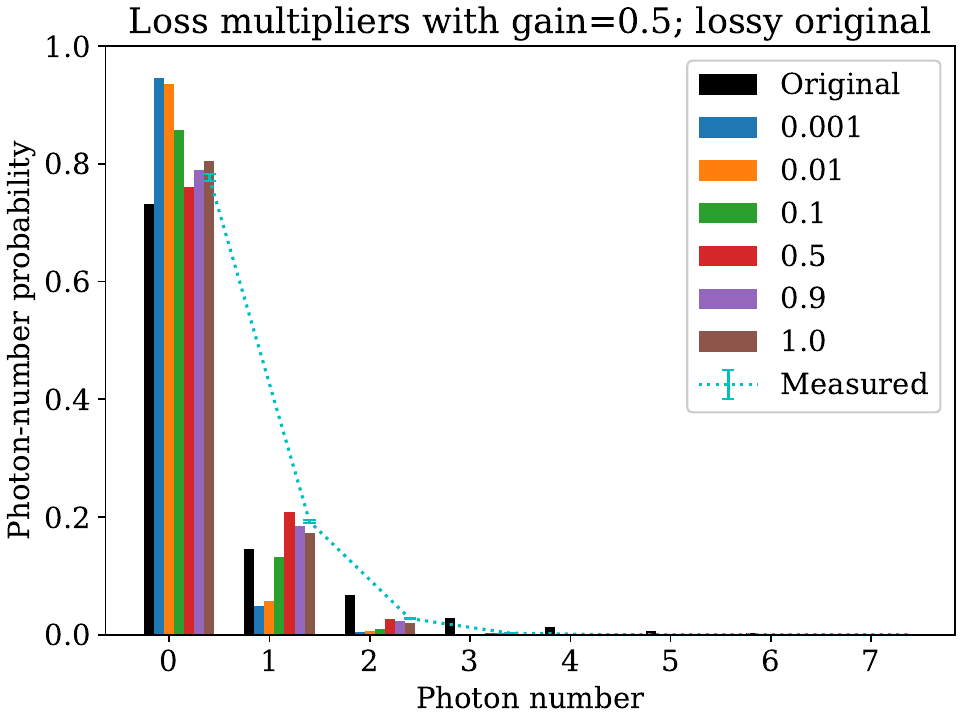}
    \caption{Same as Fig.~\ref{fig:g2 lossy orig} but with gain parameter $g=1/2$.}
    \label{fig:g05 lossy orig}
\end{figure}

\begin{figure}
    \centering
    \includegraphics[width=0.5\columnwidth]{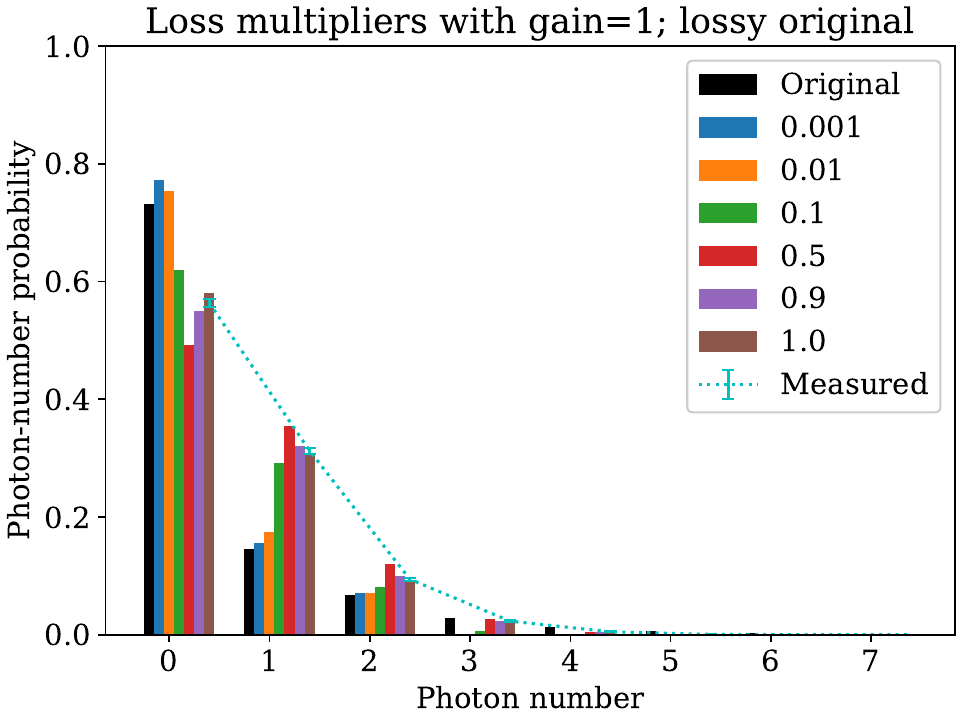}
    \caption{Same as Fig.~\ref{fig:g2 lossy orig} but with gain parameter $g=1$.}
    \label{fig:g1 lossy orig}
\end{figure}

\begin{figure}
    \centering
    \includegraphics[width=0.5\columnwidth]{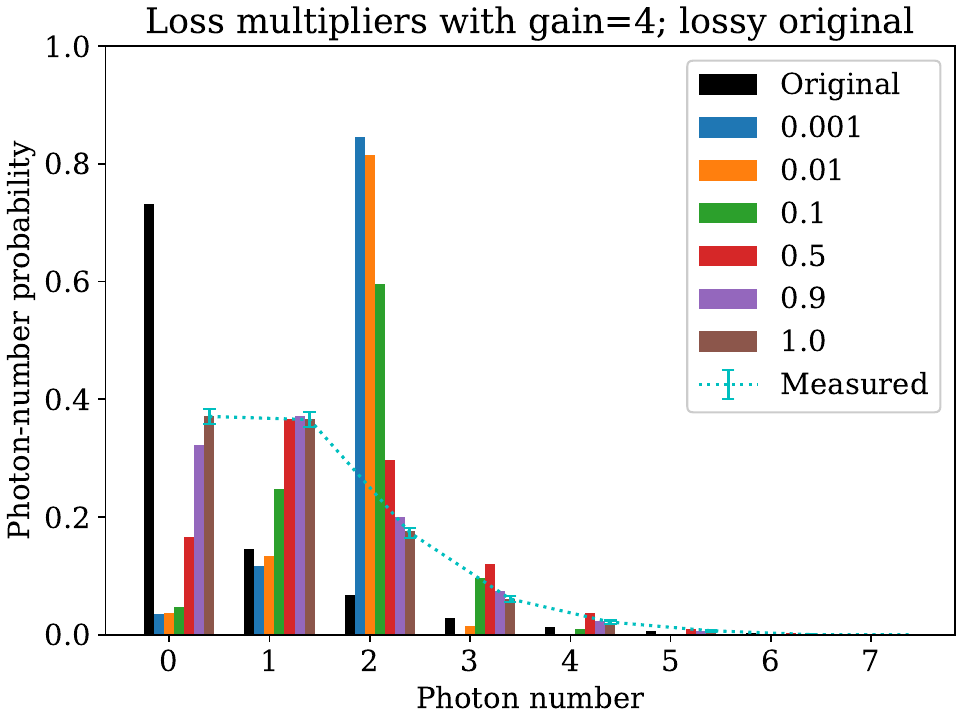}
    \caption{Same as Fig.~\ref{fig:g2 lossy orig} but with gain parameter $g=4$.}
    \label{fig:g4 lossy orig}
\end{figure}

We again consider reducing the loss by multiplicative factors, but the original state to be an attenuated SMSV that resulted from waiting in the first delay loop for 11 lossy round trips. The figures for gains of $g=1/2$, $1$ and $4$ are plotted in Figs.~\ref{fig:g05 lossy orig}, \ref{fig:g1 lossy orig}, and \ref{fig:g4 lossy orig}, respectively.

\end{document}